\documentclass[12pt,a4paper]{article}
\usepackage{amsmath,amsthm,amsfonts}

\def\req#1{(\ref{#1})}
\renewcommand{\theequation}{\arabic{equation}}
\date{ }
\author{Israel Klich}
\begin{document}
\title{Photon Green's function and the Casimir energy in a medium} \maketitle \vskip 2mm
\begin{center}

 Departments of Applied
Mathematics and Physics, \\ Technion - Israel Institute of
Technology, Haifa 32000 Israel\footnote{e-mail:
klich@tx.techion.ac.il} \\

\end{center}
\begin{abstract}
A new expansion is established for the Green's function of the
electromagnetic field in a medium with arbitrary $\epsilon$ and
$\mu$.  The obtained Born series are shown to consist of two types
of interactions - the usual terms (denoted $\cal P$) that appear
in the Lifshitz theory combined with a new kind of terms (which we
denote by $\cal Q$) associated with the changes in the
permeability of the medium. Within this framework the case of
uniform velocity of light ($\epsilon\mu={\rm const}$) is studied.
We obtain expressions for the Casimir energy density and the first
non-vanishing contribution is manipulated to a simplified form.
For (arbitrary) spherically symmetric $\mu$ we obtain a simple
expression for the electromagnetic energy density, and as an
example we obtain from it the Casimir energy of a
dielectric-diamagnetic ball. It seems that the technique presented
can be applied to a variety of problems directly, without
expanding the eigenmodes of the problem and using boundary
condition considerations.
\end{abstract}
\bigskip


\section{Introduction}

A theory describing the fluctuations of electromagnetic fields in
dielectric medium was developed long ago by Lifshitz, Pitaevskii
and others, and is now classical textbook material
\cite{LP,BG,KK}. Among other things the theory permits efficient
and general treatment of the Van der Waals interaction and
calculation of Casimir energies. This theory was developed mainly
for cases where the medium has arbitrary unhomogenous permittivity
$\epsilon(r,\omega)$ while the material considered is not
magnetoactive (with uniform $\mu$). In another classical work,
Balian and Duplantier \cite{BaDu77} derived an expansion for the
Green's function of the electromagnetic field in vacuum, but in
the presence of a conducting boundary surface. The terms in this
expansion are related to the geometry of the surface, and are
given as explicit integrals, which differ from terms encountered
in \cite{LP}. This expansion was then applied to the study of the
statistical properties of the electromagnetic eigenmodes and led
to an expansion for the Casimir energy under arbitrary conducting
boundary conditions at zero and finite temperatures \cite{BaDu78}.

    Although having no direct relation to the dielectric material problem,
for the special case where $\epsilon\mu={\rm const}$ (this
condition will be referred to as the uniform velocity of light -
UVL in short) there is such a connection. This was first apparent
in the spherical problem where the UVL scenario was first
introduced in the present context by Brevik and Kolbendsvent
\cite{BK}. The problem of a dielectric-diamagnetic ball in a
medium under the UVL condition is evaluated by expanding either
the Green's function or the appropriate characteristic functions
for the modes in terms of the diluteness parameter
$\xi=(\mu-\mu')/(\mu+\mu')$ (where $\mu$ and $\mu'$ are the
permeabilities inside and outside the ball respectively). It was
shown that one can calculate the term proportional to $\xi^2$
\cite{Kl} exactly by means of an integral over the free Green's
function of the Helmholtz equation. While introduced as a formal
trick, it coincides exactly with the two scattering result in
\cite{BaDu78} up to multiplication by $\xi^2$. This leads us to
develop a general expansion for the Green's function in
dielectric-diamagnetic media.

    The expansion that will be presented coincides with the approach
of Lifshitz when $\mu$ is constant, and bears strong formal
similarity to the multiple scattering expansion when considering
dielectric bodies immersed in a medium in which UVL holds. It is
interesting to point out that while the parameter encountered in
many calculations in the literature of the Green's function in the
latter case is $\xi$, the natural parameter appearing in the
present expansion is $\log[(1-\xi)/(1+\xi)]$.

    We start by deriving a Born type series for the Green's
function in a medium with arbitrary $\epsilon$ and $\mu$. The
series are shown to consist of two types of interactions - the
usual terms (which will be denoted $\cal P$) that appear in the
Lifshitz theory combined with a new kind of terms (which we denote
by $\cal Q$) associated with the changes in the permeability of
the medium. The $\cal P$ terms are proportional to
$(1-\epsilon\mu)$ and thus, when UVL holds only $\cal Q$ terms
contribute to the Green's function.
    Next we investigate the properties of the $\cal Q$ terms are
investigated and it is shown that in the UVL case the magnetic and
electric Green's functions are closely related - a feature which
is not present in general settings of dielectric medium. We
proceed to use the Green's function for the calculation of the
Casimir energy of the medium.
    We study in detail the contribution of the second term in the
expansion to the energy density of the electromagnetic field, and
obtain an integral formula. For the case of an ${\it arbitrary}$
radially dependent $\mu$ this contribution can be considerably
simplified. As an example, we use the radial formula to calculate
the Casimir energy of a dielectric-diamagnetic ball and show that
the exact energy to order $\xi^2$ is easily recovered.

\section{Born type expansion}
The statistical properties of the electromagnetic field in a
medium are described by the appropriate photonic Green's function.
The electromagnetic fields are derived from the electromagnetic
potentials $A^{\alpha}$, $\alpha=0,..,3$. (It is convenient to
work in the gauge $A^0=0$.) The Green's function is now defined
by:
\begin{equation}
{\mathcal D}_{ik}(X_1,X_2)=i<TA_i(X_1)A_k(X_2)>,
\end{equation}
where $X_1,X_2$ are 4-vectors $(X_1^0,..X_1^3)$ and $i,k=1,..3$.
The angular brackets denote averaging with respect to the Gibbs
distribution and $T$ denotes the time ordering operator. In
general we will be interested in the retarded Green's function,
$D_{ik}$, which is defined by:
\begin{eqnarray}
D_{ik}(X_1,X_2)=\Big{\{}\begin{array}{ll}
i<A_i(X_1)A_k(X_2)-A_k(X_2)A_i(X_1)> \,\,\,& t_1<t_2 \\  0  & {\rm
otherwise}
\end{array}
\end{eqnarray}
This function, when Fourier transformed with respect to the time
difference is actually the generalized susceptibility of the
system \cite{LP,BG}. It is known that in a medium with a given
permittivity tensor $\epsilon_{ij}$ and permeability tensor
$\mu_{ij}$, $D$ satisfies the equation:
\begin{equation}
{\bf \nabla}\times\mu^{-1}{\bf \nabla}\times D +{\omega^2\over
c^2}\epsilon D=-4\pi\hbar {\rm I}\delta({\bf r}-{\bf r}')
\end{equation}
where ${\rm I}$ is the 3-dimensional unit matrix. In what follows
we shall work in units where $c=\hbar=1$.

It will be convenient to introduce the following notation, which
is suitable for some of the calculations ahead: Let $a$ be a
3-vector; we define the cross product operator with $a$ as
$[a]_{ik}=\epsilon_{ijk}a_j$. This operator acts on vector valued
functions by\footnote{The convenience of this notation will arise
from the necessity to evaluate expressions such as $Tr(a\times
b\times c\times)$. Written in this form one has to keep track of
the order of the multiplication because of the nonassociative
nature of vector products. Working simply with matrices such as
$[a]$ will prove to be very transparent.}
$[a]\overrightarrow{f}=\overrightarrow{a}\times\overrightarrow{f}$.
Thus, for example, the curl operator is denoted $[{\bf \nabla}]$.
Furthermore we adopt the convention that the same notation is used
for the operators and their kernels. With this notation the
equation for the Green's function (multiplied by $4\pi$) of
dielectric media has the following form:
\begin{equation}\label{MainEquation}
([{\bf \nabla}]\mu^{-1}[{\bf
\nabla}]-\omega^2\epsilon)D=-{\bf{\mathbb I }}
\end{equation}
In the following we assume that $\mu({\bf r})$ is a scalar
function (of course, in the general case both $\mu$ and $\epsilon$
are tensors).

To write this equation in a manner convenient for the following we
multiply by $\mu$ and use the fact that $\mu[{\bf
\nabla}]\mu^{-1}=-[{\bf \nabla}\log\mu]+[{\bf \nabla}]$. Thus
obtaining:
\begin{equation}\label{eqDs}
([{\bf \nabla}]^2-[{\bf \nabla}\log\mu][{\bf
\nabla}]-\omega^2\epsilon\mu)D=-\mu{\bf{\mathbb I}}
\end{equation}
The inverse of the operator $([{\bf \nabla}]^2-\omega^2)$ is well
known and will be denoted $D_0$. The kernel of $D_0$, which is the
Green's function in vacuum, is given by the formula:
\begin{equation}\label{D00}
D_0(\omega;r,r')=-\left({\mathbb I }+{1\over\omega^2}{\bf
\nabla}\otimes{\bf \nabla}\right)g_0({\bf r}-{\bf r}'),
\end{equation}
where $g_0$ is the Green's function for the Helmholtz equation:
\begin{equation}\label{Helmholtz}
\triangle g_0+\omega^2 g_0=-\delta({\bf r}-{\bf r}'),
\end{equation}
given by:
\begin{equation}
g_0(r-r')={1\over 4\pi|{\bf r}-{\bf r}'|}\exp^{(i\omega |{\bf
r}-{\bf r}'|)}.
\end{equation}
Now we wish to use the known $D_0$ to express $D$ via a Born type
series. Denote
 $P={\mathcal P}+{\mathcal Q}$ where:
\begin{equation}
{\mathcal P}=\omega^2{\bf{\mathbb I }}(1-\mu\epsilon)
\end{equation}
and
\begin{equation}
{\mathcal Q}=-[{\bf \nabla}\log(\mu)][{\bf \nabla}]
\end{equation}
Then:
\begin{equation}
D=((I- P D_0){D_0}^{-1})^{-1}\mu=D_0(I-P D_0)^{-1}\mu
\end{equation}
Thus we get the following formal series for $D$:
\begin{equation}\label{expansion}
D=D_0\mu+D_0({\mathcal P}+{\mathcal Q})D_0\mu+D_0({\mathcal
P}+{\mathcal Q})D_0({\mathcal P}+{\mathcal Q})D_0\mu+....
\end{equation}
Thus the series for $D$ consist of two types of interaction. The
$\cal P$ terms are well known and can be found in \cite{LP}.
However, to our knowledge the magnetoactive terms $\cal Q$ are new
in this context and are related to the {\it changes} in the
permeability of the medium.

\section{The case of uniform velocity of light}
Uniform velocity of light in the medium implies $\epsilon\mu\equiv
{\bf\rm I}$. This considerably simplifies the expansion
\req{expansion}, since it just leaves the $Q$ scattering terms. We
are then left with the expansion:
\begin{equation}\label{epansion}
D=D_0 \mu+D_0{\mathcal Q}D_0 \mu+D_0{\mathcal Q}D_0{\mathcal Q}D_0
\mu+....
\end{equation}

\noindent In the following we are going to apply $Q$ to the free
Green's function $D_0$. For this purpose it is convenient to
introduce\footnote{taking the rotor of $D_0$ simplifies this
function considerably since we are only left with first order
derivatives of the scalar Helmholtz propagator: indeed
$\epsilon_{ijl}\partial_j\partial_l\partial_k\equiv 0$}
\begin{equation}
G_0\equiv[{\bf \nabla}] D_0=[{\bf \nabla} g_0]
\end{equation}
Thus the action of ${\mathcal Q}$ on $D_0$ is given by
\begin{equation}\label{QD0}
{\mathcal Q}D_0=-[{\bf {\bf \nabla}}\log\mu] G_0
\end{equation}
So that \req{epansion} can be written in the form:
\begin{eqnarray}\label{ex1}
D=D_0\mu-D_0[{\bf {\bf \nabla}}\log\mu]G_0\mu+... (-1)^{n-1}
D_0([{\bf {\bf \nabla}}\log\mu]G_0)^n\mu-+....
\end{eqnarray}
The terms in \req{ex1} can be written as explicit integrals using
relation \req{QD01}:
\begin{eqnarray}\label{expansion2}
& D({\bf r},{\bf r}')=\mu({\bf r}')D_0-\mu({\bf r}')\int D_0({\bf
r},{\bf r}_1)g_0'({\bf r}_1-{\bf r}')[{\bf {\bf
\nabla}}\log\mu({\bf r}_1)][{\bf u}_{r_{1}r'}]+\\ \nonumber &
\mu({\bf r}')\int D_0({\bf r},{\bf r}_1)g_0'({\bf r}_1-{\bf
r}_2)g_0'({\bf r}_2-{\bf r}')[{\bf {\bf \nabla}}\log\mu({\bf
r}_1)][{\bf u}_{r_1r_2}][{\bf {\bf \nabla}}\log\mu({\bf
r}_2)][{\bf u}_{r_2r'}]-....
\end{eqnarray}
Convergence of the series relates of course on the nature of
$\mu(r)$. For specific cases, such as of a body immersed in
dielectric-diamagnetic medium the expansion will have a natural
parameter, as will be explained in section 4.
\subsection{The magnetic Green's function and its relation to the electric Green's function}
Another quantity of interest is the magnetic Green's function,
given by: $\overrightarrow{{\bf \nabla}}\times D
\times\overleftarrow{{\bf \nabla}}$. In this section we show that
the expansion for the magnetic Green's function is similar to that
of the electric Green's function, the main difference being
alternating signs of the terms. This property is important for
calculation of the Casimir energy, since it cancels odd terms in
the expansion.

Indeed let us write
\begin{eqnarray}
D=(I+D_0[{\bf \nabla}\log\mu][{\bf \nabla}])^{-1}D_0\mu=
D_0(I+[{\bf \nabla}\log\mu]G_0)^{-1}\mu
\end{eqnarray}
Thus,
\begin{eqnarray}
& \overrightarrow{{\bf \nabla}}\times D \times\overleftarrow{{\bf
\nabla}}=\\ \nonumber & [{\bf \nabla}]([{\bf \nabla}]D^t)^t=[{\bf
\nabla}]([{\bf \nabla}]\mu D_0 (I+(D_0[{\bf \nabla}\log\mu][{\bf
\nabla}])^t)^{-1})^t=\\ \nonumber & [{\bf \nabla}](I+D_0[{\bf
\nabla}\log\mu][{\bf \nabla}])^{-1}(-D_0[{\bf
\nabla}\log\mu]\mu+[{\bf \nabla}]D_0\mu )=\\ \nonumber &
(I+G_0[{\bf \nabla}\log\mu])^{-1}(-G_0[{\bf
\nabla}\log\mu]\mu+\omega^2 D_0\mu-\mu)=\\ \nonumber & \omega^2
(I+G_0[{\bf \nabla}\log\mu])^{-1}D_0\mu -\mu
\end{eqnarray}
(Here we used the elementary property
$(1+AB)^{-1}A=A(1+BA)^{-1}$.) We can now write the expansion of
the magnetic Green's function in the form:
\begin{eqnarray}
& \overrightarrow{{\bf \nabla}}\times D \times\overleftarrow{{\bf
\nabla}}=\\ \nonumber &-\mu+\omega^2 D_0 \mu+\omega^2 (D_0[{\bf
{\bf \nabla}}\log\mu]G_0)^t\mu+... \omega^2 (D_0([{\bf {\bf
\nabla}}\log\mu]G_0)^n)^t\mu+....
\end{eqnarray}
Comparing this expansion with the one for $D$ (Eq\req{ex1}), we
see that the magnetic Green's function has the same type of
expansion as $D^t$, but with constant signs, owing to $[{\bf
\nabla}\log\mu]$ being the only antisymmetric operator in the
expansion. A similar symmetry was obtained in a different way for
the Green's function in vacuum with conducting boundary conditions
in \cite{BaDu77} and its implication on the distribution of
eigenmodes is discussed. The implication of this symmetry to the
calculation of the Casimir energy, is that in the series for the
energy the main term one has to consider is the second order term
in the expansion. This follows since the zeroth term is usually
subtracted as the free problem solution while the first term
cancels between the electric contribution and the magnetic
contribution.

\subsection{The symmetry when relaxing the UVL condition}

What can be said when keeping, in addition, the $\mathcal P$
terms? Can we still have cancellation or similarity between the
electric and magnetic Green's function? Cancellation in the spirit
of the preceding section can be achieved by expanding as follows:
Let $D_{\epsilon}$ be the solution of
\begin{equation}
([{\bf \nabla}][{\bf
\nabla}]-\omega^2\epsilon\mu)D_{\epsilon}=-{\bf{\mathbb I }}
\end{equation}
And accordingly $G_{\epsilon}=[{\bf \nabla}]D_{\epsilon}$. Then it
is possible to expand $D$ in terms of $D_{\epsilon}$. Using the
fact that $D_{\epsilon}$ is a solution for a {\it non
magnetoactive} problem and therefore has the symmetry property
$D_{ik}({\bf r},{\bf r}')=D_{ki}({\bf r}',{\bf r})$. One may show
that a there is a again a symmetry between the expansions of the
magnetic and electric Green's functions, similarly to the one
discussed in the previous section.

\section{Dielectric-diamagnetic body immersed in a medium}
In this section we consider a body ${\bf B}$ with constant
dielectric properties $\epsilon,\mu$ immersed in a medium with
different dielectric properties $\epsilon',\mu'$. Let $\partial
{\bf B}$ denote the boundary of the body, and let $s$ be
coordinates on this boundary. In this case the ${\mathcal Q}$
tensor, as a function of its first variable, will be supported on
the boundary of the body (more generally, on $\partial {\bf
B}\times \mathbb{R}^3$) and has the following structure:
\begin{equation}
{\mathcal Q}({\bf s},{\bf r})=\log({\mu\over \mu'})[{\bf n(s)}]
G_0({\bf s},{\bf r})
\end{equation}
where ${\bf n(s)}$ is the unit normal to $\partial {\bf B}$ at $
{\bf s}$.

\noindent When UVL holds, the expansion for $D$ is:
\begin{eqnarray}\label{expansionG}
& D({\bf r},{\bf r}')=\mu({\bf r}')D_0-\log({\mu\over
\mu'})\int_{\partial {\bf B}} d{\bf s}_1 D_0({\bf r},{\bf
s}_1)[{\bf n(s_1)}] G_0({\bf s}_1,{\bf r}')+\\ \nonumber &
\log({\mu\over \mu'})^2\int_{\partial {\bf B}} d{\bf s}_1 d{\bf
s}_2 D_0({\bf r},{\bf s}_1)[{\bf n(s_1)}] G_0({\bf s}_1,{\bf
s}_2)[{\bf n(s_2)}] G_0({\bf s}_2,{\bf r}')-....
\end{eqnarray}
This can be further simplified using \req{expansion2}:
\begin{eqnarray}\label{expansion3}
& D({\bf r},{\bf r}')=\mu({\bf r}')D_0-\log({\mu\over
\mu'})\mu({\bf r}')\int_{\partial {\bf B}} d{\bf s}_1 D_0({\bf
r},{\bf s}_1)g_0'({\bf s}-{\bf r}')[{\bf n}_s][{\bf u}_{sr'}]+\\
\nonumber & \mu({\bf r}')\log({\mu\over \mu'})^2\int_{\partial
{\bf B}} d{\bf s}_1 d{\bf s}_2 D_0({\bf r},{\bf s}_1)g_0'({\bf
s}_1-{\bf s}_2)g_0'({\bf s}_2-{\bf r}')[{\bf n}_{s_1}][{\bf
u}_{s_1s_2}][{\bf n}_{s_2}][{\bf u}_{s_2r'}]-....
\end{eqnarray}
Thus the Green's function is expanded in terms of the parameter:
$\log({\mu\over \mu'})$. It is interesting to relate this
parameter to the parameter $\xi$, which appears in the literature
\cite{BK} when considering the case of dielectric-diamagnetic ball
or cylinder. To see the relation we expand:
\begin{equation}\label{logmutoxi}
\log{\mu\over \mu'}=\log{1-\xi\over 1+\xi}=-2\xi-2{\xi^3\over
3}+...
\end{equation}
Hence the two expansions can now be related by assuming $\xi$
small. In particular the second term, proportional to
$(\log{\mu\over \mu'})^2$ is exactly one forth of the coefficient
of $\xi^2$ in the common expansions of the Casimir energy.
\section{Energy density from Green's function}
Fluctuation dissipation theory says that (with our definition of
$D$), at temperature $T={1\over \beta}$ the correlation functions
of the fields $A_i$ are related to the generalized susceptibility
$D$ by \cite{LP}:
\begin{equation}
(A_i({\bf r})A_j({\bf r}'))_{\omega}=-i4\pi\coth({\beta\omega\over
2})( D_{ij}({\bf r},{\bf r}')-\overline{D_{ji}}({\bf r}',{\bf r}))
\end{equation}
Now one can check that
\begin{equation}
(E_iE_k)_{\omega}=\omega^2(A_iA_k)_{\omega}
\end{equation}  and that
\begin{equation}
(B_iB_k)_{\omega}=\overrightarrow{{\bf \nabla}}\times
(A_iA_k)_{\omega}\times\overleftarrow{{\bf \nabla}}.
\end{equation}
Note that there is no minus sign before the correlator of the
electric fields, this can be verified by returning to the time
parameter and using $E=-\dot{A}$. Thus the energy density per
d$\omega$ of the electromagnetic field at temperature $T$ is given
by
\begin{equation}\label{rhoT}
\rho_{T}({\bf r},\omega)=-\coth({\beta\omega\over 2})\rho({\bf
r},\omega)
\end{equation}
where
\begin{equation}
\rho({\bf r},\omega)={1\over 4\pi}{\rm Limit}_{{\bf r}'\rightarrow
{\bf r}}{\bf\rm Im}{\rm Tr} [\omega^2\epsilon({\bf r})D({\bf
r},{\bf r}')+{1\over\mu({\bf r})}\overrightarrow{{\bf
\nabla}}\times D \times\overleftarrow{{\bf \nabla}}]
\end{equation}
This expression for $\rho$ is usually divergent and should be
regularized. The natural regularization scheme in this problem
would be to subtract the energy density of the spatially
homogeneous problem for this result.
\section{Expansion for energy density and density of modes in a
medium with UVL}

As was previously shown, in the UVL case there is a symmetry
between the electric and magnetic Green's functions. As a result
the calculations are considerably simplified. In the expression
for the energy density the zeroth order terms are cancelled by the
vacuum $D_0$ function, and moreover we are left solely with the
even terms of the expansion. All the higher order terms are given
by explicit integrals in which the limit ${\bf r}'\rightarrow {\bf
r}$ can be taken at the outset. Thus, for example, the first term
to be considered in the energy density is:
\begin{equation}
{\rho_T}^{(2)}({\bf r},\omega)= -\coth({\beta\omega\over
2})\rho^{(2)}({\bf r},\omega)
\end{equation}
Where
\begin{eqnarray}
& \rho^{(2)}({\bf r},\omega)= {1\over 2\pi}\omega^2{\rm Im} \int
g_0'({\bf r}_1-{\bf r}_2)g_0'({\bf r}_2-{\bf r})\\ \nonumber
&\times{\rm Tr}(\left[{\bf {\bf \nabla}}\log\mu({\bf r}_1)][{\bf
u}_{r_1r_2}][{\bf {\bf \nabla}}\log\mu({\bf r}_2)]G_0({\bf
r}_2,{\bf r})D_0({\bf r},{\bf r}_1)\right)
\end{eqnarray}
To find the density of the Casimir energy with respect to $\omega$
this expression is to be integrated over the volume considered. It
is convenient to integrate first over ${\bf r}$ (see
\cite{BaDu77}). In our notation:
\begin{equation}
\int{\rm d}{\bf r}\,G_0({\bf r}_2,{\bf r})D_0({\bf r},{\bf
r}_1)=[{\bf \nabla}_{r_2}](I+{1\over\omega^2} {\bf
\nabla}_{r_1}\otimes{\bf \nabla}_{r_1})\int{\rm d}{\bf
r}\,g_0({\bf r}_2-{\bf r})g_0({\bf r}-{\bf r}_1)
\end{equation}
Using the identity:
\begin{equation}
\int{\rm d}{\bf r}\,g_0({\bf r}_2-{\bf r})g_0({\bf r}-{\bf
r}_1)={1\over 2\omega}\partial_{\omega}g_0 ({\bf r}_2-{\bf r}_1)
\end{equation}
We obtain the result:
\begin{equation}
\int{\rm d}{\bf r}\,G_0({\bf r}_2,{\bf r})D_0({\bf r},{\bf
r}_1)={1\over 2\omega}\partial_{\omega}G_0({\bf r}_2-{\bf r}_1)
\end{equation}
Namely we have the operator identity, $G_0 D_0={1\over
2\omega}\partial_{\omega}G_0$. This can be done at any order in
the perturbation series so that we obtain the following general
expression:
\begin{eqnarray}\label{densityexp}
& \rho^{(2n)}(\omega)= {1\over 2\pi}\omega^2 {\rm Im} \int {\rm
Tr}(\left[{\bf{\bf \nabla}}\log\mu({\bf r}_1)][{\bf
u}_{r_1r_2}]\cdot\cdot\cdot[{\bf {\bf \nabla}}\log\mu({\bf
r}_{2n})][{\bf u}_{r_{2n}r_1}]\right) \\ \nonumber &\times
g_0'({\bf r}_1-{\bf r}_2)\cdot\cdot\cdot g_0'({\bf r}_{2n-1}-{\bf
r}_{2n}){1\over 2\omega}\partial_{\omega}g_0'({\bf r}_{2n}-{\bf
r}_1)
\end{eqnarray}
In principle the Casimir energy can be obtained from these
expression to any order and for any $\mu$. However, in most cases
this is impossible to carry out, and one has to be content with a
first few orders.

\section{Casimir energy: second order results}
In this section we focus on the second term in the expansion for
the density \req{densityexp}. We will obtain a simplified
expression for this term, however, one must bear in mind that for
any specific problem it is necessary to check whether this term is
indeed larger then the other terms in the expansion. To cast the
second term in a more convenient form, it is useful to introduce
the function:
\begin{equation}
g_2({\bf r}_1-{\bf r}_2)=g_0'({\bf r}_1-{\bf r}_2){1\over
2\omega}\partial_{\omega}g_0'({\bf r}_{2}-{\bf r}_1)={e^{2i\omega
|{\bf r}_{2}-{\bf r}_1|}\over 32\pi^2|{\bf r}_{2}-{\bf
r}_1|}({1\over|{\bf r}_{2}-{\bf r}_1|}-i\omega),
\end{equation}
so that the second order contribution to the energy is given by
\req{densityexp}:
\begin{eqnarray}
\rho^{(2)}(\omega)= {1\over 2\pi}\omega^2 {\rm Im} \int {\rm
Tr}(\left[{\bf{\bf \nabla}}\log\mu({\bf r}_1)][{\bf
u}_{r_1r_2}][{\bf {\bf \nabla}}\log\mu({\bf r}_{2})][{\bf
u}_{r_{2}r_1}]\right)g_2({\bf r}_1-{\bf r}_2)
\end{eqnarray}
Using the relation \req{TwoTrace}
\begin{equation}
Tr([{\bf a}][{\bf b}][{\bf c}][{\bf b}])=2({\bf a\cdot b})({\bf
b\cdot c})
\end{equation}
We obtain
\begin{eqnarray}\label{SecondRho}
\rho^{(2)}(\omega)= {1\over 2\pi}\omega^2 {\rm Im} \int ({\bf{\bf
\nabla}}\log\mu({\bf r}_1)\cdot {\bf u}_{r_1r_2})({\bf {\bf
\nabla}}\log\mu({\bf r}_{2})\cdot {\bf u}_{r_{2}r_1})g_2({\bf
r}_1-{\bf r}_2)
\end{eqnarray}
This is our final result for the first contribution to the Casimir
energy for arbitrary $\mu$. For simple geometries this can be
still simplified. For example, if we assume that $\mu$ is a
function of one coordinate only, namely $\mu=\mu(z)$, Then $[{\bf
\nabla}\log\mu]=(\log\mu(z))'[{\bf \hat{z}}]$ and the second term
is
\begin{eqnarray}\label{so}
{\rho_T}^{(2)}(\omega)=-{1\over
2\pi}\omega^2\coth({\beta\omega\over 2}) {\rm Im} \int 2({\bf
\hat{z}}\cdot {\bf u}_{r_1 r_2})^2 g_2({\bf r}_{2}-{\bf
r}_1)(\log\mu(z_1))'(\log\mu(z_2))'
\end{eqnarray}
The energy density per unit area for this configuration can be now
obtained by integrating this expression over ${\bf r_2}$ and the
$z$ coordinate of ${\bf r_1}$.
\subsection{Second order term for spherically symmetric $\mu$}

For cases with high symmetry the equation \req{SecondRho} for the
energy density can be further simplified. Indeed, let us assume
that we are dealing with a medium with spherical symmetry. This
type of mediums can be encountered in a variety of problems from
QCD bag models to cosmology. In our formalism this amounts to
consider a magnetic permeability $\mu$ which is a radial function.
Thus $[{\bf \nabla}\log\mu]=(\log\mu)'[{\bf u}_{0r}]$. The second
term \req{SecondRho} is then given by
\begin{eqnarray}
& \rho^{(2)}(\omega)= -{\omega^2\over 2\pi} {\rm Im} \int {\rm
Tr}(\left[{\bf u}_{0 r_1}][{\bf u}_{r_1 r_2}][{\bf u}_{0
r_2}][{\bf u}_{r_2 r_1}]\right)
\\ \nonumber &\times g_2({\bf r}_1-{\bf r}_2)(\log\mu(|{\bf
r}_{1}|))'(\log\mu(|{\bf r}_{2}|))'
\\ \nonumber &=-{\omega^2\over 2\pi} {\rm Im} \int
2({\bf u}_{0 r_1}\cdot {\bf u}_{r_1 r_2})({\bf u}_{0 r_2}\cdot
{\bf u}_{r_1 r_2}) \\ \nonumber &\times g_2({\bf r}_1-{\bf
r}_2)(\log\mu(|{\bf r}_{1}|))'(\log\mu(|{\bf r}_{2}|))'
\end{eqnarray}
To carry the integration further, we change to bipolar coordinates
$u=|{\bf r}_{2}-{\bf r}_1|$ and $v=|{\bf r}_{2}|$ and $r=|{\bf
r}_{1}|$. Thus:
\begin{equation}
({\bf u}_{0 r_1}\cdot {\bf u}_{r_1 r_2})({\bf u}_{0 r_2}\cdot {\bf
u}_{r_1 r_2})={(v^2-u^2-r^2)\over 2ru}{(r^2-v^2-u^2)\over
2vu}={(u^4-(v^2-r^2)^2)\over 4u^2rv}
\end{equation}
and the integration becomes:
\begin{eqnarray}
& {\rm Im} \int_{0}^{\infty}4\pi r^2{\rm dr}(\log\mu(r))' \int
{2\pi uv\over r}{\rm dudv} {(u^4-(v^2-r^2)^2)\over 2u^2rv}
g_2(u)(\log\mu(v))'\\ \nonumber &
\end{eqnarray}
Where the integration domain is such that $u$, $v$ and $r$ can
form a triangle. More specifically, the integrations are given by:
\begin{eqnarray}
I(r)\equiv \left(\int_{0}^{r}{\rm du}\int_{r-u}^{r+u} {\rm dv}+
\int_{r}^{\infty}{\rm du} \int_{u-r}^{u+r} {\rm dv}\right)\left(
{(u^4-(v^2-r^2)^2)\over 2u} g_2(u)(\log\mu(v))'\right)
\end{eqnarray}
Finally, the second order contribution for the Casimir energy in
this case is given by the formula:
\begin{eqnarray}\label{radialmu}
& {\rho_T}^{(2)}(\omega)= -4\pi\omega^2\coth({\beta\omega\over 2})
{\rm Im} \int_{0}^{\infty}{\rm dr}I(r)(\log\mu(r))'
\end{eqnarray}
This equation is our final formula for the Casimir energy of a UVL
configuration with arbitrary radially symmetric $\mu$. As an
example, in the next subsection we implement it for a one-line
calculation of the Casimir energy of a dielectric diamagnetic
ball.
\subsection{Example: The dielectric-diamagnetic ball}
It was first shown by Boyer \cite{Bo68}, that, contrary to
intuition, the Casimir force on a conducting spherical shell is
repulsive. A large number of subsequent calculations by various
techniques \cite{aB,BaDu78} verified Boyer's result. In the
following, we show that the calculation of the Casimir energy of a
ball of radius $a$ immersed in a medium, under the UVL, condition
is immediate using the technique developed above. In this case
$\mu$ is of the form: $\mu(r)=\mu$ if $r<a$ and $\mu(r)=\mu'$ if
$r>a$. Let $\kappa=\log{\mu\over\mu'}$. Then from \req{radialmu}
it is immediate that the density of states is of the form
\begin{equation}\label{rhoball}
{\rho_T}^{(2)}(\omega)= -4\pi\omega^2\coth({\beta\omega\over 2})
{\rm Im} (\kappa\, I(a))
\end{equation}
Now explicitly calculating $I(a)$ we have:
\begin{equation}\label{II}
I(a)=\kappa\int_0^{2a}{u^3\over 2} g_2(u)
\end{equation}
Performing the integral in \req{II} and substituting in
\req{rhoball}, we have:
\begin{equation}\label{RhoBall}
\rho^{(2)}(\omega)=-{\rm Im}\left[{-1+e^{4ia\omega} -
4ie^{4ia\omega}a\omega-4e^{4ia\omega}\omega^2 a^2\over {8\pi
}}\right]\xi^2+{\mathcal O}(\xi^4)
\end{equation}
(Where we used \req{logmutoxi} to write the result in terms of
$\xi$.) Taking the imaginary part of the expression on the right
gives one a simple way of eliminating the divergence in this
problem, which arises upon integrating over frequencies. This
simply eliminates the term proportional to $1/8\pi$ in the energy
density. Alternatively, this term is independent of radius and
thus can be eliminated by subtracting the unconstrained system
where $a\rightarrow\infty$. In any case we recover the result
obtained in \cite{Kl} by mode summation.

\noindent At zero temperature, the integration over frequencies is
conveniently carried out on the imaginary axis to yield the known
result to order $\xi^2$:
\begin{equation}
E=\int_{-\infty}^{\infty}{d\omega
\frac{\xi^2}{8\pi}e^{-4a|\omega|}(1+4a|\omega|+4a|\omega|^2)}=\frac{\xi^2}{2
a}\frac{5}{16 \pi}
\end{equation}
Since the energy is a positive quantity decreasing with radius the
Casimir pressure on the boundary will be outward. At finite
temperatures, one has to take into account the
$\coth({\beta\omega\over 2})$ factors that appear in \req{rhoT}.
This is conveniently done by changing the integration to summation
over the imaginary Matsubara frequencies. In \cite{KlFe} this was
done using the expression \req{RhoBall}, with the result
:
\begin{eqnarray}\label{ballenergy}
& E_C(a,T)=\frac{T\xi^2}{4}+\frac{T\xi^2}{2\left( e^{8a\pi
T}-1\right)}+\frac{4a\pi T^2\xi^2e^{8a\pi T}} {{{\left(  e^{8a\pi
T}-1\right) }^2} }+ \\ \nonumber &
\frac{8a^2\pi^2{T^3}\xi^2e^{8a\pi T}\left(1+e^{8a\pi T}
\right)}{{\left(  e^{8a\pi T}-1\right) }^3}
\end{eqnarray}
This result is exact (to order $\xi^2$) for all temperatures. It
is easy to check that the pressure on the boundary remains outward
at all temperatures.

\section{Concluding remarks}
 In this paper we presented a technique appropriate
for evaluation of the Green's function of dielectric media,
especially in cases where uniform velocity of light (UVL) holds.
The perturbative technique developed allows one to deal with the
Casimir effect for geometries not studied in other approaches. It
was shown that for the UVL case the magnetic and electric Green's
functions are closely related - a feature which is not present in
general settings of dielectric medium. Due to this property the
first contribution to the energy density of the electromagnetic
field comes from the second term in the expansion. We have shown
that equation \req{SecondRho} is an explicit integral
representation of this contribution. For cases of spherical
symmetry we derived a simplified formula, namely \req{radialmu},
for the energy density of a configuration with an arbitrary radial
$\mu$. From this formula the energy of a dielectric-diamagnetic
ball follows easily, without recourse to expansion of the modes
themselves or boundary condition considerations, other problems
can now be also accessed in the same manner.

\section*{Acknowledgments}
I wish to thank L.P. Pitaevskii for helpful discussions. I also
wish to thank A. Elgart, A. Mann, and M. Revzen for many useful
remarks. \noindent

\section*{Appendix: Identities}
For the convenience of the reader we list here a few simple
relations:
\setcounter{equation}{0}
\renewcommand{\theequation}{A.\arabic{equation}}

\noindent {\bf 1.}
\begin{equation}
[{\bf b}][{\bf a}]={\bf a\otimes b-a\cdot b{\bf{\mathbb I }}}
\end{equation}
This is easy to see by working with indices:
\begin{equation}
([{\bf b}][{\bf a}])_{im}=\epsilon_{ijk}b_j\epsilon_{klm}a_l=
(\delta_{il}\delta_{jm}-\delta_{im}\delta_{jl})a_lb_j=a_i
b_m-\delta_{im}a_lb_l
\end{equation}
{\bf 2.}
\begin{equation}
G_0({\bf r},{\bf r}')=g_0'({\bf r}-{\bf r}')[{\bf u}_{rr'}]
\end{equation}
where ${\bf u}_{rr'}$ is a unit vector in the direction from ${\bf
r}$ to ${\bf r}'$. Indeed:
\begin{equation}
G_0({\bf r},{\bf r}')=[{\bf \nabla}]D_0=-[{\bf \nabla}] g_0({\bf
r}-{\bf r}'){\bf{\mathbb I }}=-[{\bf \nabla}_{({\bf r}-{\bf r}')}
g_0({\bf r}-{\bf r}')]=g_0'({\bf r}-{\bf r}')[{\bf u}_{rr'}]
\end{equation}
{\bf 3.} From these two we conclude that:
\begin{equation}\label{QD01}
{\mathcal Q}D_0=[{\bf {\bf \nabla}}\log\mu({\bf r})] G_0({\bf
r},{\bf r}')=g_0'({\bf r}-{\bf r}')[{\bf {\bf \nabla}}\log\mu({\bf
r})][{\bf u}_{rr'}]
\end{equation}
{\bf 4.} Symmetry properties
\begin{eqnarray}
& (i)\,\,{D_0}_{ij}({\bf r},{\bf r}')={D_0}_{ji}({\bf r}',{\bf
r})\,\,\,\,\,\,\,\,\,\,(ii)\,\,{G_0}_{ij}({\bf r},{\bf r}')=
{G_0}_{ji}({\bf r}',{\bf r})\\ \nonumber &
(iii)\,\,{G_0}_{ij}({\bf r},{\bf r}')=-{G_0}_{ji}({\bf r},{\bf
r}')=-{G_0}_{ij}({\bf r}',{\bf r})\,\,\,\,\,\,\,\,\,\,
(iv)\,\,[{\bf a}]=-[{\bf a}]^t
\end{eqnarray}
{\it Remark:} Property $(i)$ is true in the form $D_{ij}({\bf
r},{\bf r}')=\overline{D_{ji}({\bf r}',{\bf r})}$ for any $D$
which is a susceptibility function for non-magnetoactive medium
\cite{LP}.

\noindent {\bf 5.}
\begin{equation}
A\times\overleftarrow{{\bf \nabla}}=([{\bf \nabla}]A^t)^t
\end{equation}
{\bf 6.}
\begin{equation}
\overrightarrow{{\bf \nabla}}\times D_0 \times\overleftarrow{{\bf
\nabla}'}=\omega^2D_0-{\bf{\mathbb I }}
\end{equation}
Indeed:
\begin{eqnarray}\label{cdc}
\overrightarrow{{\bf \nabla}}\times D_0 \times\overleftarrow{{\bf
\nabla}'}=[{\bf \nabla}]([{\bf \nabla}] (D_0)^t)^t=[{\bf
\nabla}]G_0=[{\bf \nabla}][{\bf \nabla}]D_0=-{\bf{\mathbb I
}}+\omega^2 D_0
\end{eqnarray}
\noindent {\bf 7.} Using the relations $({\bf a\otimes b})({\bf
c\otimes d})=({\bf b\cdot c}){\bf a\otimes d}$ and ${\rm Tr}({\bf
a\otimes b})=({\bf a\cdot b})$ we get
\begin{eqnarray}\label{TwoTrace}
& {\rm Tr}(\left[{\bf u}_{0 r_1}][{\bf u}_{r_1 r_2}][{\bf u}_{0
r_2}][{\bf u}_{r_1 r_2 }]\right)= \\ \nonumber & {\rm Tr}(\left
({\bf u}_{0 r_1}\otimes {\bf u}_{r_1 r_2}-{\bf u}_{0 r_1}\cdot
{\bf u}_{r_1 r_2}{\bf{\mathbb I }})({\bf u}_{0 r_2}\otimes {\bf
u}_{r_1 r_2}-{\bf u}_{0 r_2}\cdot {\bf u}_{r_1 r_2}{\bf{\mathbb I
}})\right)=
\\ \nonumber & 2({\bf u}_{0 r_1}\cdot {\bf u}_{r_1 r_2})({\bf u}_{0 r_2}\cdot {\bf
u}_{r_1 r_2})
\end{eqnarray}


\end{document}